\documentclass[twoside,11pt]{article}

\usepackage{jmlr2e}
\usepackage{xspace}
\usepackage{pifont}

\newcommand{\lib}{{\texttt{dagger}}\xspace}

\newcommand\digitstyle{\color{purple}}

\usepackage{listings}
\usepackage{xcolor}
 
\definecolor{codegreen}{rgb}{0,0.6,0}
\definecolor{codegray}{rgb}{0.5,0.5,0.5}
\definecolor{codepurple}{rgb}{0.58,0,0.82}
\definecolor{backcolour}{rgb}{0.95,0.95,0.92}
 
\lstdefinestyle{pystyle}{
    backgroundcolor=\color{backcolour},   
    commentstyle=\color{codegreen},
    keywordstyle=\color{magenta},
    numberstyle=\tiny\color{codegray},
    stringstyle=\color{codepurple},
    basicstyle=\ttfamily\tiny,
    emph={and,break,class,continue,def,yield,del,elif ,else,%
        except,exec,finally,for,from,global,if,import,in,%
        lambda,not,or,pass,print,raise,return,try,while,assert,with,as},
    emphstyle=\color{magenta}\bfseries,
    emph={[2]True,False},
    emphstyle=[2]\color{orange}\bfseries,
    breakatwhitespace=false,         
    breaklines=true,                 
    captionpos=b,                    
    keepspaces=true,                 
    numbers=left,    
    numbersep=2pt,
    showspaces=false,                
    showstringspaces=false,
    showtabs=false,                  
    tabsize=2, 
    otherkeywords = {@}
}
 
\makeatletter

\AtBeginDocument{%
  \newcounter{llabel}[lstlisting]%
  \renewcommand*{\thellabel}{%
    \ifnum\value{llabel}<0 %
      \@ctrerr
    \else
      \ifnum\value{llabel}>10 %
        \@ctrerr
      \else
        \protect\ding{\the\numexpr\value{llabel}+201\relax}%
      \fi
    \fi
  }%
}
\newlength{\llabelsep}
\newcommand*{\llabel}[1]{%
  \begingroup
    \refstepcounter{llabel}%
    \label{#1}%
    \llap{%
      \thellabel\kern\llabelsep
      \hphantom{\lst@numberstyle\the\lst@lineno}%
      \kern\lst@numbersep
    }%
  \endgroup
}
\newcommand{\ProcessDigit}[1]
{%
  \ifnum\lst@mode=\lst@Pmode\relax%
   {\digitstyle #1}%
  \else
    #1%
  \fi
}

\makeatother
\ShortHeadings{A Framework for Reproducible Machine Learning Experiment Orchestration}{Paganini and Forde.}
\firstpageno{1}

\begin{document}

\title{\lib: A Python Framework for Reproducible Machine Learning Experiment Orchestration}

\author{\name Michela Paganini \email michela@fb.com \\
       \addr Facebook AI Research \\
       \AND
       \name Jessica Z. Forde \email jzf2101@columbia.edu \\
       \addr Brown University, Facebook AI Research
       }

\editor{TBD}

\maketitle

\begin{abstract}
Many research directions in machine learning, particularly in deep learning, involve complex, multi-stage experiments, commonly involving state-mutating operations acting on models along multiple paths of execution. Although machine learning frameworks provide clean interfaces for defining model architectures and unbranched flows, burden is often placed on the researcher to track \textit{experimental provenance}, that is, the state tree that leads to a final model configuration and result in a multi-stage experiment. Originally motivated by analysis reproducibility in the context of neural network pruning research, where multi-stage experiment pipelines are common, we present \lib, a framework to facilitate reproducible and reusable experiment orchestration. We describe the design principles of the framework and example usage.
\end{abstract}


\section{Introduction}

Experiment tracking and cataloguing, as well as the recording of systematic data and model provenance, are key to building confidence and trust in the scientific process.

In the context of machine learning research, the lack of reproducibility standards and the complexity of experimental pipelines have given rise to a known reproducibility crisis, and to ample conversation and contributions to try to address it~\citep{sonnenburg2007need}.
However, even in the case of open-sourced code releases, it is common practice to share incomplete ad-hoc experimental boilerplate, often fused with the core technical contribution and lacking experiment configuration specifications, making solutions harder to disentangle from the infrastructure, and build upon. 

This work addresses the need for a lightweight, modular, \textit{model-centric} machine learning workflow-creation solution that allows researchers to abstract away fundamental scientific contributions from experiment-tracking boilerplate code, while drawing causal inheritance relations among model states in a fully reproducible manner. \lib is a minimal framework for describing trees of network-mutating actions suited to the needs of researchers, allowing fast experimentation as well as maintenance of clear provenance in experiment evolution.

\lib is made available under the MIT License, and is accessible on GitHub\footnote{\texttt{https://github.com/facebookresearch/dagger}}. \lib is tested with Continuous Integration from CircleCI for Linux and MacOS platforms.

\section{Related Work}
The need and desire to track complex, evolving state in an immutable graph structure is not new -- such a treatment of objects, and operations on such objects are quite common in the orchestration~\citep{netflix-conductor, uber-cadence}, workflow~\citep{airflow, luigi}, and data processing~\citep{spark, dask} communities.
Although comprehensive, such frameworks are often best suited to cluster-level production usage, and often introduce a high-surface-area interface, which makes them unsuitable for fast experimentation, a requirement of researchers. 
Another family of open-source solutions addresses experiment management~\citep{klaus_greff-proc-scipy-2017}, but scope is often limited to hyperparameter tracking. By comparison, \lib puts model evolution first, providing the primitives to analyze model changes over time (and allowing for hyperparameter management as a subcase).

\section{Design}

\subsection{Formalism}
\label{ssec:formalism}
To motivate the need for such a framework, consider the following experimental setup. Let $S \subset\mathcal{S}$ represent a \emph{state} (most commonly representing a model configuration and its context, \textit{e.g.} training hyperparameters, data set, random seed, etc.). We define any transformation $R : \mathcal{S} \longrightarrow \mathcal{S}$ as a \emph{recipe}, that is, a manner by which to mutate state. Examples of state-mutating actions include, but are not limited to, model training, initialization, pruning, and quantization. Graphically speaking, we represent any specific transition $R_j: S_i \mapsto S_{j}$ between a pair of nodes $S_i$ and $S_j$ as an edge $(S_i, S_{j})$ with edge value of $R_j$. We require the existence of a root state, $S_0$, and require acyclicity and connectedness in the graph $G=(V,E,W)$ defined by nodes $V=\{S_i\}$, edges $E=\{(S_i, S_j)\}$, and edge values $W=\{R_j\}$, affording users the ability to track the provenance and unique path from any state $S_j\leadsto S_0$. Note that $\vert E \vert = \vert W \vert = N - 2$, by the definition of a tree.

\subsection{Implementation}
\label{ssec:implementation}
The entities defined in Sec.~\ref{ssec:formalism} map cleanly onto the library surface area. The outermost entity that \lib provides is an \texttt{Experiment} object, which allows users to lazily define the experiment graph for later execution. An experiment is located in a directory on the file system, in which all states are serialized to facilitate caching.
The \texttt{ExperimentState} class represents a node unit in the experiment tree, and provides bookkeeping and hashing capabilities.
Users are expected to customize state definition by subclassing the \texttt{ExperimentState} class and overriding the \texttt{PROPERTIES} and \texttt{NONHASHED\_ATTRIBUTES} class attributes. 
To separate definition from execution, \lib internally uses an \texttt{ExperimentStatePromise} object, which symbolically represents a future \texttt{ExperimentState}. 

By subclassing the \texttt{Recipe} object and defining a \texttt{run} method, users specify and bound the set of custom actions that, according to the logic of the experiment, cause a state to mutate into a new child state, \textit{i.e.} a new node in the graph.

Finally, non-state-mutating actions, such as model performance evaluation, which does not modify the state nor its context, are supported via the \texttt{Function} class.


The definition of states and recipes allows full caching of the computational graph. For a state $S_j$ with parent\footnote{We allow for a parent to be null, for the unique case of the root state.} $S_i$, such that $S_j = R_j(S_i)$, we can (recursively) compute a hash as $h_j = H(S_j, h_i)$, for a suitably chosen function $H$, where $h_i$ is the hash of the parent. This avoids duplicate computation when attaching new ops to a preexisting experiment tree.

\lib is built on Dask~\citep{dask} for the underlying lazy evaluation infrastructure, and, as a result, can run in single-threaded (default), multi-threaded, multi-process, and distributed environments. Since \lib's aim is for a broad set of machine learning researchers to use opinionated experiment orchestration to promote reproducibility, the framework is deep learning-library agnostic, cross-platform, and hardware-independent.

\subsection{Example Usage}
\label{ssec:usage}
\begin{lstlisting}[label={experiment_def}, language={Python}, caption={Example code for custom experiment object and action definition.}, escapechar=|]
import dagger as dg
from yourlib import get_data, get_model, train_model, prune_model, eval_model

class State(dg.ExperimentState):

|\llabel{properties}|    PROPERTIES = ["dataset_name", "model_name"]
|\llabel{nonhashed}|    NONHASHED_ATTRIBUTES = ["train_data", "eval_data", "model"]

|\llabel{init_state}|    def initialize_state(self, **kwargs):
        self.train_data, self.eval_data = get_data(self.dataset_name)
        self.model = get_model(self.model_name)

class TrainRecipe(dg.Recipe):

|\llabel{recipe_properties}|    PROPERTIES = ["nb_epochs", "lr"]

|\llabel{recipe_run}|    def run(self, state):
        train_model(state.model, state.train_data, self.nb_epochs, self.lr)
        return state
        
class PruneRecipe(dg.Recipe):

    PROPERTIES = ["pruning_technique", "pruning_fraction"]

    def run(self, state):
        prune_model(state.model, self.pruning_technique, self.pruning_fraction)
        return state

@dg.function
|\llabel{functions}|def eval_fn(state):
    eval_acc = eval_model(state.model, state.eval_data)
    print(f"Experiment: {state.tags}, Accuracy: {eval_acc}")
\end{lstlisting}

\begin{lstlisting}[label={experiment_run}, language={Python}, caption={Example code for a realistic training and pruning experiment run.}, escapechar=|]
exp = dg.Experiment("/path/to/experiment/folder", state_class=State)
|\llabel{spawn}|root_state = exp.spawn_new_tree(dataset_name="cifar-10", model_name="vgg-11")

for lr in [0.01, 0.1]:
    train = TrainRecipe(nb_epochs=100, lr=lr)
    prune = PruneRecipe(pruning_technique="lowest_magnitude", pruning_fraction=0.2)
    s = root_state
|\llabel{tags}|    with exp.tag(f"lr:{lr}"):
        s = train(s)
        eval_fn(s)
        with exp.tag("pruned"):
            s = prune(s)
|\llabel{run}|exp.run()
\end{lstlisting}

This section outlines the usage of \lib in a simple, illustrative scenario where we compare two pruned models obtained from trainings with different learning rates.

In Listing~\ref{experiment_def}, experiment setup takes place. Within the user-defined \texttt{State} class, \ref{properties} and \ref{nonhashed} allow users to define the properties of a state as well as the instance attributes that child states inherit from parent states. The \texttt{initialize\_state} method \ref{init_state} is a required override in the subclass that defines any initialization beyond simple assignment\footnote{As examples, this can cover initializing models, getting data, setting seeds, and detecting and setting desired compute hardware.}, and is called inside \texttt{spawn\_new\_tree} (see \ref{spawn} in Listing~\ref{experiment_run}). The subclassed \texttt{TrainRecipe} defines properties~\ref{recipe_properties}, used to develop the state's hash $h$, as well as the \texttt{run} method~\ref{recipe_run}, used to define the state transition operation.  \lib also provides support for functions~\ref{functions}, which do not add to the graph, but execute in-graph, even when the graph is cached.

\begin{figure}[!h]
    \centering
    \includegraphics[width=\textwidth]{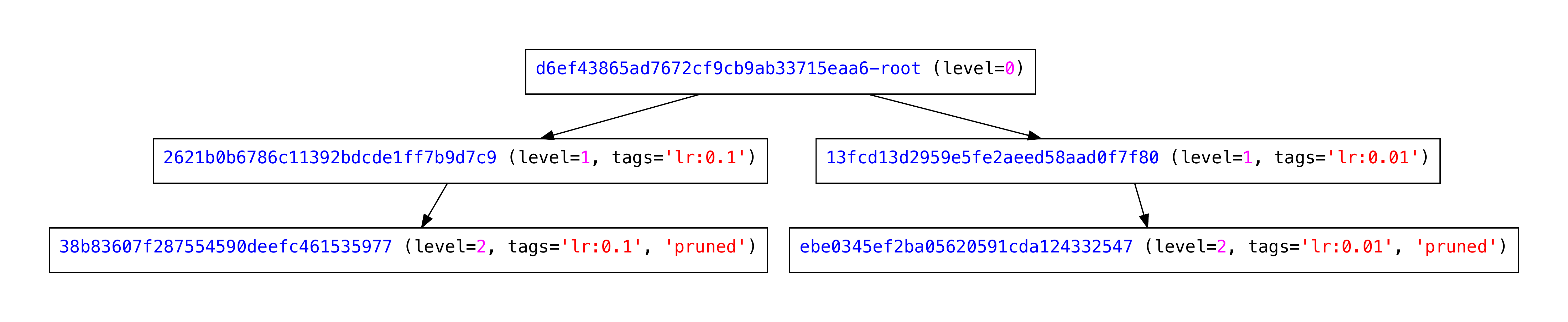}
    \caption{Visualized Graph from \texttt{Experiment} in Listing~\ref{experiment_run} obtained by call to \texttt{exp.graph.draw()} in Listing~\ref{visualize}.
    }
    \label{fig:graph}
\end{figure}

In Listing~\ref{experiment_run}, the experiment tree is defined and run. Most rapid research iterations will take place here.
When a new \texttt{Experiment} is started, the \texttt{spawn\_new\_tree} method~\ref{spawn} can be used to instantiate a root \texttt{ExperimentState}, whence all other states originate. The \texttt{Experiment} object is used to lazily keep track of the tree. Individual states can have tags added~\ref{tags}, which facilitate experiment analysis (see \ref{filter} in Listing~\ref{visualize}) and visualization (Figure~\ref{fig:graph}). \texttt{Recipe}s handle the creation of descendent nodes. The graph can be run on a single core, or can be scaled to a cluster~\ref{run}.

The analysis API (Listing~\ref{visualize}) affords the ability to load a \texttt{slim} version of the graph in memory, filter and isolate specific \texttt{ExperimentState}s~\ref{filter}, and restore those states in memory for specific usage or analysis~\ref{restore}.

\begin{lstlisting}[label={visualize}, language=Python, caption={Loading and filtering an experiment graph via the analysis API.}, escapechar=|]
>>> exp = Experiment.restore("/path/to/experiment/folder", slim=True)
>>> exp.graph.draw()  # Draws the graph in Figure 1
|\llabel{filter}|>>> s = exp.graph.nodes.filter("pruned") & exp.graph.nodes.filter("lr:0.1")
|\llabel{restore}|>>> s[0].restore()
\end{lstlisting}

\newpage

\bibliography{sample}

\end{document}